**Mitigating E-beam-induced Hydrocarbon Deposition on Graphene for Atomic-Scale Scanning Transmission Electron Microscopy Studies**

# Mitigating E-beam-induced Hydrocarbon Deposition on Graphene for Atomic-Scale Scanning Transmission Electron Microscopy Studies

Ondrej Dyck[a)], Songkil Kim, Sergei V. Kalinin, and Stephen Jesse

The Center for Nanophase Materials Sciences, Oak Ridge National Laboratory, Oak Ridge, Tennessee, 37831, USA

The Institute for Functional Imaging of Materials, Oak Ridge National Laboratory, Oak Ridge, Tennessee, 37831, USA

[a)]Electronic mail: dyckoe@ornl.gov

CVD grown graphene used in (scanning) transmission electron microscopy ((S)TEM) studies must undergo a careful transfer of the one-atom-thick membrane from the growth surface (typically a Cu foil) to the TEM grid. During this transfer process, the graphene invariably becomes contaminated with foreign material. This contamination proves to be very problematic in the (S)TEM because often >95% of the graphene is obscured and imaging of the pristine areas results in e-beam-induced hydrocarbon deposition which further acts to obscure the desired imaging area. In this article, we examine two cleaning techniques for CVD grown graphene that mitigate both aspects of the contamination problem: visible contamination covering the graphene, and "invisible" contamination that deposits onto the graphene under e-beam irradiation. The visible contamination may be removed quickly by a rapid thermal annealing to 1200 ºC in situ and the invisible e-beam-deposited contamination may be removed through an Ar/$O_2$ annealing procedure prior to imaging in the (S)TEM.

## I. INTRODUCTION

Atomically resolved imaging via scanning probe and electron microscopy has opened the doors to the nanoworld by providing a pathway to visualize atomic structure and explore functional properties on a single atom and molecule level.[1] Multiple examples of these studies include the order parameter fields in ferroic materials,[2, 3] octahedra tilts[4-7] and chemical strains,[8] and local chemical properties[9, 10] in (scanning) transmission electron microscopy (STEM) and energy loss spectroscopy (EELS). In the field of scanning probes (SPM), the examples include the superconductive order parameter mapping,[11] protein unfolding spectroscopies,[12, 13] exotic Kitaev phases[14] and many others.

However, both for (S)TEM and SPM, the success of atomically resolved imaging often hinges on the availability of the well-prepared samples. For (S)TEM, typically the requirements include the stability and ability to form thin foils, whereas the surface stability on the order of 1-2 nm is often irrelevant. In comparison, SPM studies are critically dependent on the stability of the top surface layer, and as a result the SPM studies are often centered on material that can be prepared by cleaving (layered materials), or sputtering and annealing and related techniques (metals, semiconductors). The number of SPM studies of materials that require more complex sample preparation such as in-situ pulsed laser deposition[15-22] or magnetron sputtering[23] growth is much more limited.

The requirements for the sample preparation are becoming much more stringent on transition from imaging to fabrication. In STM, more than 25 years was required to transition from the first atomic manipulation experiments by Don Eigler[24] to the single-atom device fabrication by M. Simmons[25-27] and others. Nowadays, it is recognized that STEM can also be used as a tool for single atom fabrication, where the electron beam is employed to induce controllable chemical transformation including vacancy,[28, 29] ad-atom,[30] and interstitial motion,[31, 32] bond formation,[31, 33-35] vacancy ordering,[36] phase changes,[37] etc.[38, 39] that can further be atomically resolved. Combination of STEM with controlled beam motion and image-based feedback further enables atom-by-atom fabrication in STEM.

Similar to SPM, the fabrication of atomic structures via STEM necessitates high-quality sample preparation. In many cases, chemical vapor deposition (CVD) grown graphene samples are prepared for (S)TEM investigation using a poly(methyl-methacrylate) (PMMA) -mediated approach for transfer from the growth substrate to the TEM grid.[40-42] With this technique, after the graphene is grown on a metal foil substrate, it is mechanically stabilized with a coating of PMMA, after which the metal foil is etched away. The graphene, attached to the layer of PMMA, may then be transferred to an arbitrary substrate and the PMMA removed with solvents. Though this approach appears to be quite common, it leaves behind PMMA and other organic residue so that the graphene is almost completely obscured from view when examined in a (S)TEM (see Figure 1). In addition, samples prepared in this way also exhibit significant e-beam induced hydrocarbon deposition, so that the small areas of pristine graphene will become covered in mobile contaminants upon imaging and thus obscured. Many attempts have been made to address these issues with varied success and they generally involve a thermal annealing of some kind. Van Dorp et. al.[43] show that heating exfoliated few-layer graphene to 500 ºC for 10 minutes in the microscope is sufficient to remove the contaminant materials. These results are supported by the investigations of Xie et. al.[44] who performed X-ray photoelectron spectroscopy (XPS) and time-of-flight secondary ion mass spectrometry (ToF-SIMS). Lin et. al.[45, 46] advance an air and $H_2$/Ar 200 °C annealing procedure prior to examination in the TEM, which appears to be quite effective. Liang et. al.[47] introduce an alternate PMMA transfer method that appears to clean graphene but they fail to demonstrate cleanliness at the atomic level. Finally, Li et. al.[48] give evidence that exposing clean CVD-grown graphene to air results in significant hydrocarbon contamination within just a few minutes. This may result in conflicting reports where graphene that has undergone a cleaning treatment, is no longer clean when investigated in the (S)TEM and the cleaning treatment is assumed to have failed. What none of these studies address is the e-beam-deposited hydrocarbon contamination that typically occurs under high magnification in the STEM.

Here we investigate two methods for cleaning graphene to address both visible contaminant materials present on the graphene surface and particularly focus on contamination resulting from e-beam-induced hydrocarbon deposition and discuss our observations. The first method we investigate is an ex situ Ar/$O_2$ anneal suggested by Garcia et. al..[49] The second method we investigate is in situ annealing.

## II. EXPERIMENTAL

CVD-grown graphene was transferred from the Cu foil growth substrate to a TEM sample grid followed by an Ar/$O_2$ anneal for removal of volatile adsorbents. The Cu foil was spin-coated with PMMA to stabilize the graphene and the Cu foil was etched away in a bath of ammonium persulfate-DI water solution. The graphene/PMMA layer was transferred to a DI water bath to remove residues of ammonium persulfate. The graphene was transferred to the final TEM substrate by scooping it from the bath and letting it dry at room temperature. TEM samples were baked in an oven under an Ar/O$_2$ (90%/10%) environment to remove residual PMMA and volatile organic compounds.

For the in-situ heating experiments graphene was transferred to a Protochips Aduro heating holder and heated at a rate of 1000 ºC/ms in the microscope.

Imaging of the samples was performed in a Nion UltraSTEM U100 at an accelerating voltage of 60 kV in high angle annular dark field (HAADF) imaging mode. The samples were loaded into the microscope using our standard loading procedure, where the microscope magazine, cartridges, and samples are baked in a vacuum chamber at 160 ºC for eight hours prior to insertion into the microscope, except where explicitly indicated in the text (i.e. discussion related to the results presented in Figure 5 (a, b, c)).

## III. RESULTS AND DISCUSSION

### A. Control Sample

To compare the cleaning techniques to a standard sample, images were taken of a graphene sample prior to any cleaning. Figure 1 shows an as deposited sample at a variety of magnification levels, exhibiting heavy contamination. The observable clean areas are extremely small and not even visible at the lowest magnification shown. Even when these small, clean areas are found, hydrocarbon e-beam deposition is often an intractable problem which acts to quickly cover the region of interest with amorphous carbon. This problem can be mitigated to some extent by performing a so-called "beam shower" by illuminating a large area for ~30 minutes. It is thought that this procedure deposits volatile contamination onto the sample across the exposed area leaving the vacuum slightly cleaner. In our experience, this technique appears to work to some degree for a period of time, sometimes as long as 1-2 hours, before it must be repeated. Nevertheless, this situation is not ideal and here we explore other options.

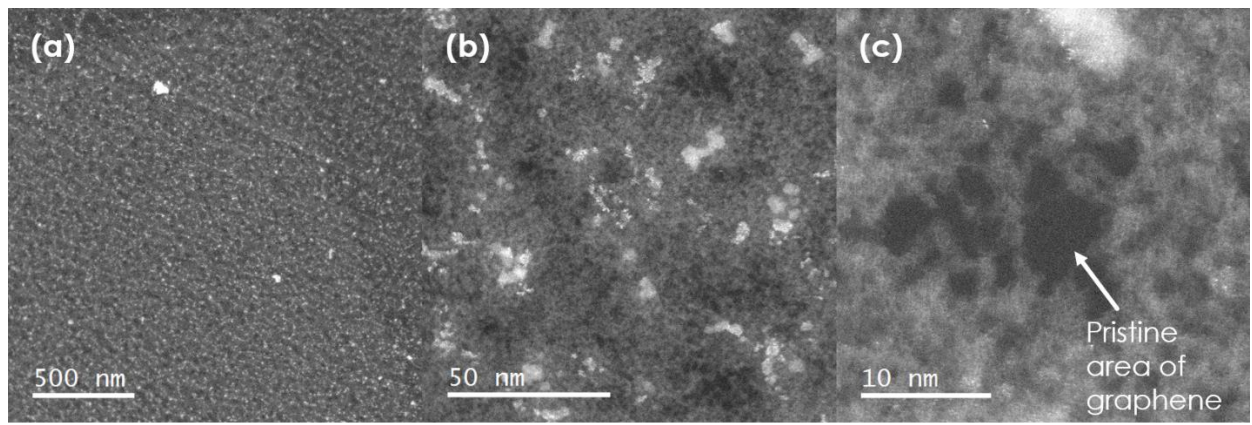

**Figure 1** (a, b, c) example images at a variety of magnification levels of an as-transferred graphene sample prior to cleaning. In (c) we can see a small patch of clean graphene about 5 nm wide, indicated by the arrow.

## B.  Ar/$O_2$ Annealing

To produce more favorable graphene samples for STEM studies that exhibit no e-beam hydrocarbon deposition we adopted the cleaning procedure of Garcia et. al..[49] This procedure involves heating the graphene sample to 500 °C in an environment of Ar (90%) and $O_2$ (10%). The studies of Garcia et. al. investigated sample cleanliness through the use of raman spectroscopy and were primarily targeted at removing adhesive residue from exfoliated h-BN. We applied this technique to CVD grown graphene, transferred to a TEM grid through a PMMA-stabilized transfer process as described in the experimental section.

Figure 2 shows the contamination morphology after the Ar/$O_2$ cleaning procedure. Figures (a-d) show the observed morphology at a variety of magnifications after microscope alignment. The central area in each image displays pristine areas of graphene due to exposure to the electron beam during alignment. These areas appear dark as they are only a single atom thick. Initially, there appears to be a significant amount of contamination covering the graphene (the areas away from the center). It is unclear, quantitatively, whether it is truly cleaner than the as-transferred graphene as far as contamination coverage is concerned because the contamination appears to contract on itself or move away from the illuminated area under e-beam influence. In other words, simply observing the sample acts to change it. Lattice resolved images produce large (10-20 nm) clean areas of graphene within a few seconds which were not present previously. To illustrate this phenomenon on a mesoscopic scale, a large area was selected and illuminated in parallel (not scanned) for 20-30 mins. The before and after images are shown in (e, f), where we observe a significant increase in pristine graphene areas over the entire illumination area (dashed circles). The insets in (e, f) show a zoomed-in portion of the same area before and after illumination. We can see that the exposed contamination is revealing pristine graphene similar to the central portion of (b). Additional images and a video time-lapse series of this phenomenon are provided in the supporting materials. Of note: smaller areas scanned with the beam become clean more quickly so that, after microscope alignment, imaging and experimentation can begin immediately with ample areas of pristine graphene.

While this result was rather unexpected and fortuitous, the truly remarkable property of such samples (remarkable based on our previous experience with graphene samples) is that we have not observed any e-beam induced hydrocarbon deposition *at any magnification* on samples cleaned with this method. Lattice resolved images may be taken at leisure for many hours without fear of encroaching contamination. Indeed, the contamination contracts away from the irradiated area, exposing more pristine graphene. This is important, as graphene sample quality has historically made atomic level studies difficult and sometimes impossible. The mechanisms of motion and chemical make-up of the contamination observed are beyond the scope of this article. We simply wish to highlight this recipe as a highly effective way to produce atomically clean areas of CVD grown graphene for (S)TEM studies.

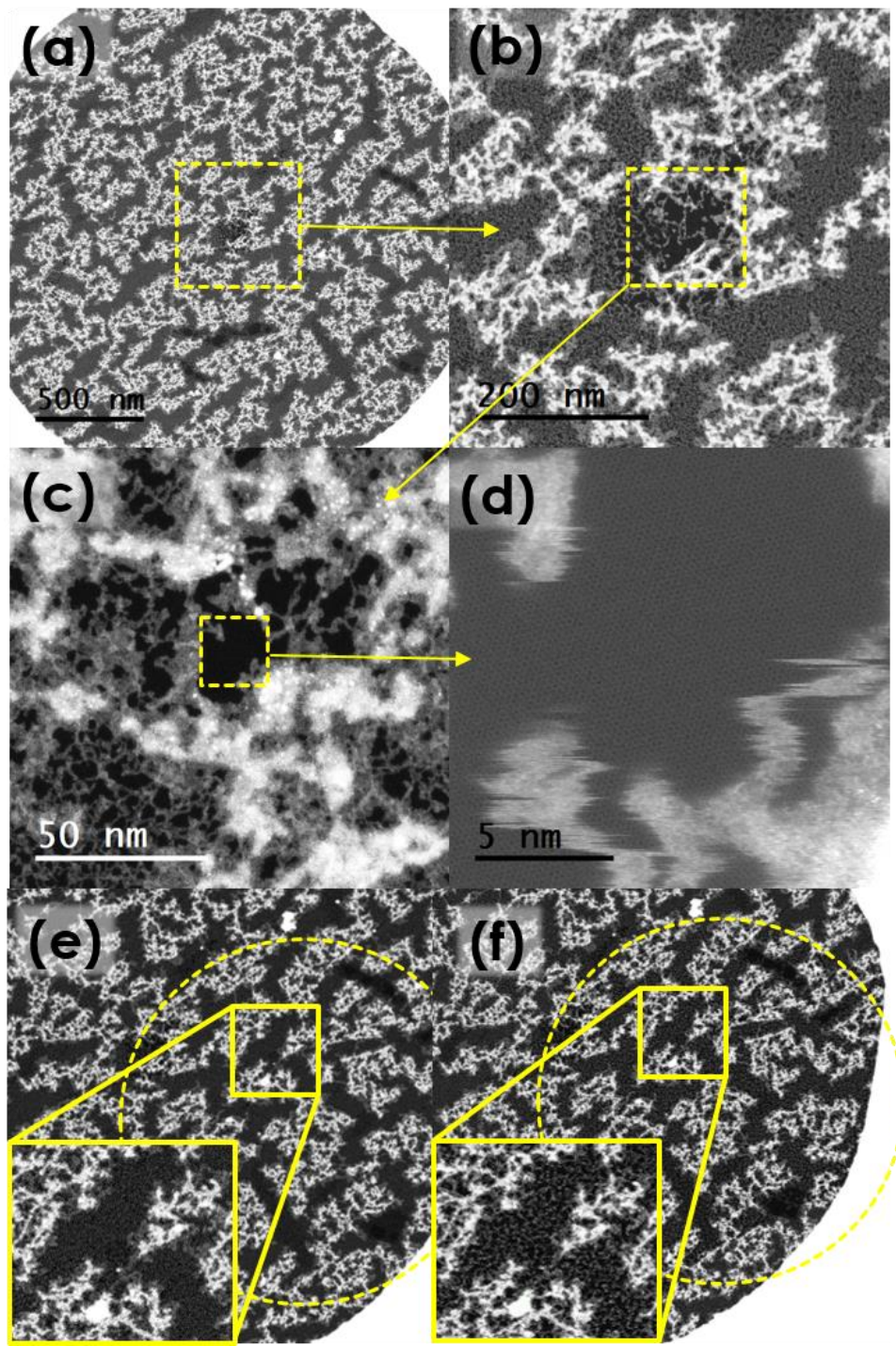

**Figure 2** (a-d) show a series of HAADF images acquired at a variety of magnifications showing the mesoscopic and atomic-level contaminant morphology. The boxed areas indicate the acquisition location of each subsequent image. In (d) we observe significant streaking of the surface contamination indicating high mobility under e-beam irradiation. As short video clip of this motion is included in the supplementary information. In (e, f) we show the result of parallel e-beam illumination on the contaminant morphology. The image in (e) was acquired before significant e-beam irradiation. The large circle indicates the approximate illumination area and the inset shows a zoomed-in portion of the same image to show more clearly the smaller features. In (f) we show the results after e-beam illumination over the area indicated by the circle. The inset shows the same area shown inset in (e). We observe a significant increase of contamination-free areas (i.e. the darkest regions).

## *C. In-Situ Rapid Thermal Cleaning*

While the above described method for cleaning graphene produces highly agreeable samples for (S)TEM studies, the samples are nevertheless, still covered in contaminant material. In order to fabricate a graphene sample that is atomically clean on the micron length scale we used a Protochips Aduro heater chip to heat the sample to 1200 ºC at a rate of 1000 ºC/ms. The results are summarized in Figure 3. In (a, b, c) we see the typical contamination of an as-prepared sample at several magnifications. After heating the sample, the images in (d, e, f) were acquired and we see that the suspended graphene film is mostly atomically clean graphene over the entire observable window. The location that remained dirtiest is boxed in (d, e) and corresponds to the previously irradiated area, discussed later. This rapid thermal cleaning procedure was repeated with several different samples with similar results.

Several observations are noteworthy:

1) After returning the sample to room temperature, the sample remained clean while in the microscope.

2) After returning the sample to room temperature, high e-beam fluence, such as that produced when performing lattice resolved images or under a stationary beam, produced heavy e-beam-induced hydrocarbon deposition. This deposition is shown in Figure 4 (a, b) and may be controllably deposited with the e-beam, (b). We posit that this contamination comes from areas of the heater chip that remain cool even while the chip is heated (the heated area is concentrated to within a few tens of microns around the sample).

3) This amorphous deposited carbon contamination can be converted to graphitic carbon upon heating again to 1200 °C (see supplemental materials).

4) Heating the sample again to 500 °C still shows e-beam deposited contamination but heating to 800 °C prevents this deposition (see supplemental materials). This is important in practice because, although the heater chip has the capability to ramp to 1200 °C, we find that the sample usually exhibits a slight mechanical instability (vibration) at this temperature, reducing resolution. Backing off from this limit appears to be more mechanically stable. As a result, the lattice resolved image in Figure 3 (f) was acquired at 800 °C since lattice resolution was not possible at 1200 °C ((d, e) were acquired at 1200 °C). Though there is much to be explored here, we limit this paper to these cursory observations.

5) We note that any areas that had previously undergone e-beam irradiation or deposition do *not* become clean with the described heating procedure (see also the observations of van Dorp et. al.[43]) but can be made graphitic upon heating. Figure 4 (c) shows a suspended sheet of graphene that had been previously imaged in an SEM to check for the success of the sample transfer. The entire region remains mostly covered in contamination, save for a few smaller patches, and areas exposed to higher e-beam fluence by increasing magnification in the SEM are clearly visible as areas with higher contaminant coverage (i.e. the brighter patches in (c)). This appears to be similar to the adherence of the contamination noted in Figure 3 (d, e) which had been exposed briefly to the STEM beam before heating. Though, undesirable for cleaning graphene, this is interesting because it immediately suggests possible patterning in a lower magnification instrument (SEM, FIB, etc.) and subsequent thermal treatment to remove the unexposed materials. This technique may be amenable to producing conductive carbon nanowires on h-BN, for example.

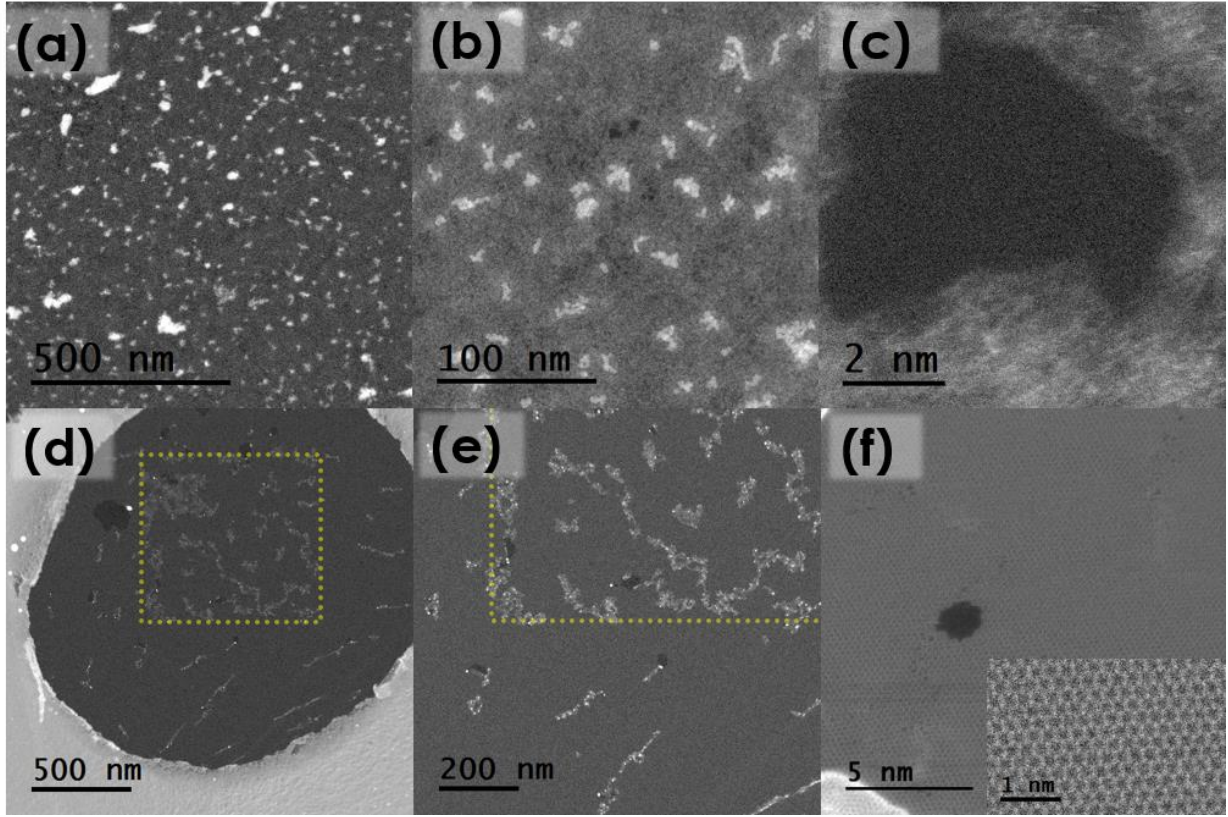

**Figure 3** (a, b, c) show the initial contamination morphology of the as-deposited sample prior to heating at three magnification levels. We see the typical contamination expected with only very small pristine areas. (d, e, f) show the suspended graphene film at a variety of magnifications after the rapid thermal annealing. We note that atomically clean graphene exists over a majority of the film extending ~2 microns. The inset in (f) is an image of a higher magnification to more clearly show the graphene lattice.

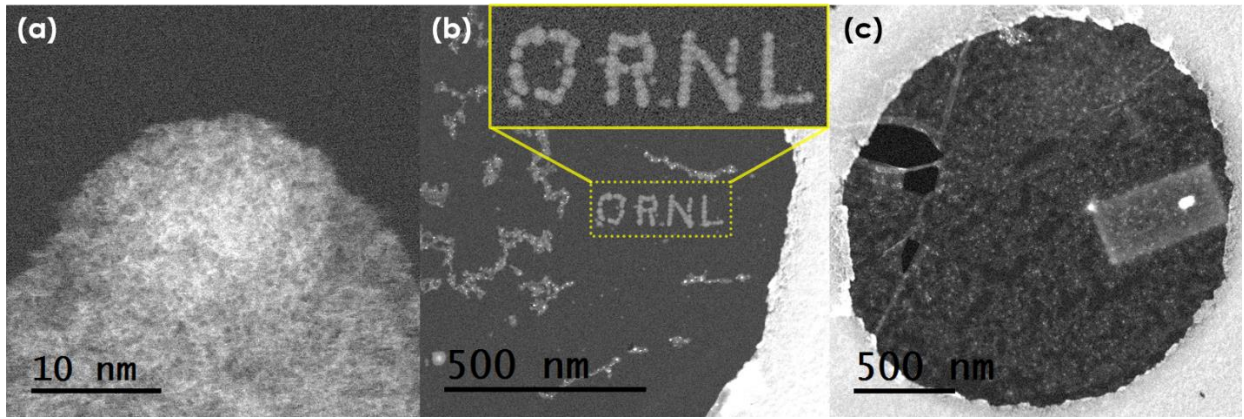

**Figure 4** (a) shows the e-beam deposited hydrocarbon contamination after brief exposure. (b) illustrates the controllable deposition of hydrocarbon contamination. Exposure to an electron beam prior to heating acts to "pin" the contamination to the sample. (c) shows an area of the sample which was exposed to an SEM beam with varying degrees of exposure prior to examination in the STEM. We see that the graphene remains highly contaminated even after heating and the greater the degree of exposure (i.e. the smaller the scan area in SEM) the more contamination remains and in the highly exposed areas appear to have additional carbon deposited on them.

Finally, we wanted to explore what happens upon removal of a clean graphene surface from the STEM vacuum and exposure to the laboratory air. We thus, cleaned a graphene sample with the above described heating procedure, cooled the sample back to room temperature, removed the sample from the microscope vacuum, and let it sit on a "gloves only" sample exchange table under a plastic (non-airtight) cover to prevent environmental dust from settling on it. After approximately four hours the sample was reintroduced into the microscope and imaged again.

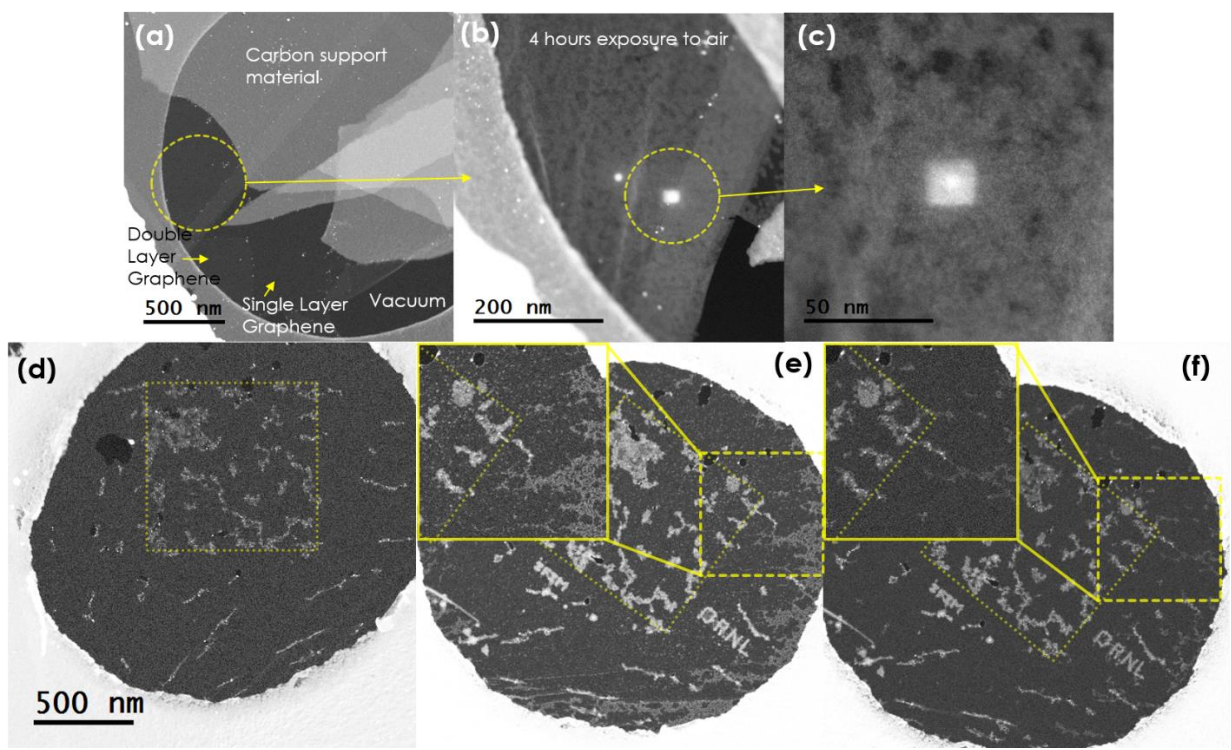

**Figure 5** (a) shows the graphene sample at 1200 °C. The structural support material for this particular sample failed upon heating and much of the sample was lost. Nevertheless, the region shown contains pristine single and multi-layer graphene, labeled. The brighter material entering the image at the top right is the carbon support material from the heater chip which covered a portion of the suspended graphene upon heating. (b) shows the circled area of the sample after four hours of air exposure and reintroduction into the STEM. We observed extremely heavy e-beam induced hydrocarbon contamination. Re-heating this sample was not possible as the microscope vacuum levels began to degrade and the heater had to be turned off. (c) shows a magnified view of the area circled in (b). (d, e, f) show the same sample featured in Figure 3 and 4 after being removed from the microscope and reintroduced the next week using the standard sample loading procedure (160 °C in vacuum for 8 hours). (d) shows the sample as it appeared after the initial in situ heating (i.e. Figure 3 (d)) with the dotted box indicating the location of previous e-beam exposure that did not become as clean. (e) shows the same suspended area after storage in air for a week. The largest differences are the presence of dendritic structures protruding from the edge onto the graphene surface as well as smaller speckles of contamination observed over the extent of the graphene. The inset shows a magnified view of the boxed area to better show fine detail. We note that the sample appears mostly clean, in contrast to (a, b, c). (f) shows a similar view as (e) after heating again to 1200 °C. Most of the dendritic structure and speckling of contamination was removed.

Figure 5 (a, b, c) show the results of this procedure. The image in (a) was acquired while the sample was being held at 1200 °C. We observe the same atomically clean graphene over extended lengths. The heater chip support material failed during heating and can be seen laying over the graphene (labeled). After removing the sample from the microscope for four hours and rintroducing it, the image shown in (b) was acquired. We observe severe e-beam deposition of hydrocarbon contamination and what appears to be a continuous layer of contamination over the entire sheet of graphene, (c) shows a magnified view of the contamination layer and deposition. Lattice resolved images were not possible due to the e-beam deposition, so it is possible there were small, nanometer-sized, areas of pristine graphene remaining. It is unclear whether the heavy contamination is sourced from the air or from adjacent areas of the sample which hadn't been heated, nevertheless we feel it is sufficient to conclude that pristine graphene is highly susceptible to the adsorption of hydrocarbon contamination in agreement with the investigations of Li et. al..[48] What is interesting, however, is that loading a sample following our standard loading proceedure, where we heat the magazene and cartridge with the sample to 160 °C for eight hours, is sufficient to remove the volitile hydrocarbon contamination observed in (b). Images of this observation are shown in (d, e, f). The sample shown in Figure 3 and 4 was removed from the microscope and stored for a week before being imaged again. We observe modest additional contamination following this prodedure. The sample still exhibited e-beam-induced hydrocarbon contamination (see supplemental materials) but it is not nearly as severe as that shown in (b, c) and visible contaminants observed on the surface were limited to a few dendritic structures that appear to have grown from the edge of the support substrate as well as speckles of contamination over the entire surface of the graphene. Upon heating again to 1200 °C, (f), most of this additional contamination was removed.

This result suggests that, although pristine graphene becomes heavily contaminated in air (Figure 5 (a, b, c)), a vast majority of this contamination may be removed by heating to 160 ºC in vacuum for 8 hours (Figure 5 (d)) which is starkly different from the kind of contamination observed in Figure 1 on the control sample which had also undergone the same heating proceedure, yet remained heavily contaminated.

## IV. SUMMARY AND CONCLUSIONS

We have explored two effective graphene cleaning procedures which we have been using to clean graphene for our STEM studies and we have provided our observations regarding their effectiveness for this purpose. In particular, we note that the $Ar/O_2$ cleaning procedure produces very agreeable samples for (S)TEM investigation at room temperature without risk of e-beam-induced hydrocarbon deposition. From a physics standpoint these samples are most interesting because there are other "contaminant" atoms sitting on the surface and frequently found in the lattice, or can even be put into the lattice, as we have recently demonstrated.[50] This provides a wealth of physical phenomenon to explore on a single sample. From an engineering perspective (building devices) such a sample may not be ideal since it is still technically covered with contamination. In order to produce atomically clean graphene on a mesoscopic scale (microns) for device fabrication, the $Ar/O_2$ cleaning procedure may be insufficient. In order address this, we demonstrated the effectiveness of a rapid thermal annealing procedure where the sample is heated to 1200 ºC at a rate of 1000 ºC/ms, which immediately[§] produces atomically clean graphene. We also detail several observations regarding what occurred at various temperatures following the cleaning, namely: returning to room temperature retains the clean graphene but significant e-beam hydrocarbon deposition is observed, reheating of the amorphous deposited carbon converts it to graphite, and heating to 800 °C is sufficient to prevent hydrocarbon deposition under the e-beam. In addition, we also observe that clean graphene readily contaminates with volatile species in air which can mostly be removed by a 160 °C anneal for eight hours in a vacuum chamber. While the observations noted here are wide ranging, and each result is not extensively investigated from a physical perspective, we hope that, from a practical perspective, these results will be of help to the microscopy community in preparing contamination free graphene samples for atomic-scale studies.

---

[§] We were unable to determine how quickly this process occurs. In the time it took to open the gun valve and examine the sample again, the graphene was clean. So, for practical purposes in (S)TEM investigations, this procedure appears to "immediately" produce clean graphene.

## SUPPLEMENTARY INFORMATION

See supplementary material at [URL will be inserted by AIP Publishing] for additional images and a video of the ArO$_2$-cleaned samples under e-beam exposure, and additional images documenting hydrocarbon deposition under various conditions.

## ACKNOWLEDGMENTS

We would like to thank Dr. Ivan Vlassiouk for provision of the graphene samples and Dr. Francois Amet for performing the argon-oxygen cleaning procedure.

Research supported by Oak Ridge National Laboratory's Center for Nanophase Materials Sciences (CNMS), which is sponsored by the Scientific User Facilities Division, Office of Basic Energy Sciences, U.S. Department of Energy (S.V.K.), and by the Laboratory Directed Research

**Supplementary Information**

In Figure S1 we show several example images of an Ar/O$_2$ cleaned sample. In a) we show how scanning the beam over the material causes the contamination to contract and reveal pristine graphene. b) and c) show higher magnification images of the pristine areas of graphene where we can see the adsorbed atoms and those occupying hole edges and substitutional positions in the lattice. d) and e) are the first and last frames of the accompanying time-lapse video of the contamination dynamics. The two images were acquired 3.7 minutes apart.

Figure S2 (a, b) show the results of heating the hydrocarbon deposition to 1200 °C. By the discrete steps in intensity we conclude that the amorphous contamination was converted to a graphitic structure. After the initial 1200 °C cleaning, (c, d, e) show a series of HAADF images, acquired in succession, illustrating the hydrocarbon deposition observed at 500 °C. This deposition occurs more gradually than that observed at room temperature and appears to be graphitic as evidenced by the steps in intensity observed in (f), which was acquired after the deposition in (c, d, e) at a lower magnification. (g, h) were acquired at 800 °C after the initial 1200 °C cleaning. E-beam deposition was no longer observed at this temperature and lattice images were able to be acquired.

Figure S3 shows what we observed after removing the cleaned sample from the microscope, storing it for a week, and reexamining it. Prior to loading the sample, it was baked in vacuum at 160 °C for eight hours. The most notable change observed was the presence of dendritic carbonaceous material extending from the edges of the support membrane.

This result is in stark contrast to that observed in Figure 5 (b, c) where excessive hydrocarbon contamination was observed to cover the graphene. We thus conclude that the mild heat treatment (160 °C for eight hours in vacuum) is sufficient to remove most of this type of contamination from graphene. Nevertheless, higher beam fluence occurring from smaller scan areas like those shown in Figure S3 (b, c, d) exhibited hydrocarbon deposition, as expected. Heating the sample again to 1200 °C removes much of the dendritic structure, as shown in Figure S3 (f, g).

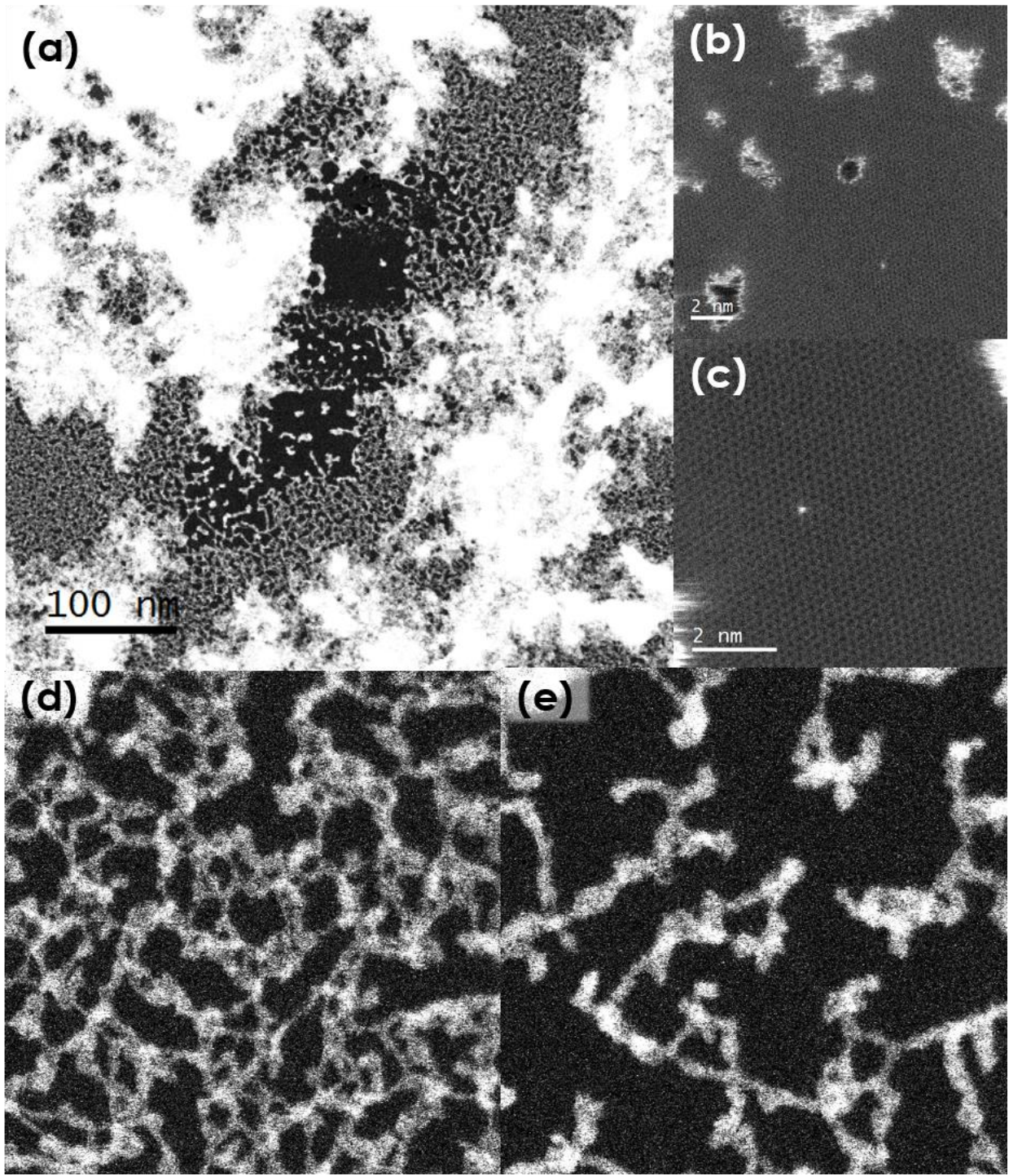

**Figure S1** (a, b, c) show a variety of magnifications of the Ar/O$_2$ cleaned sample after the e-beam was scanned over several locations. In particular, in (a) we can clearly distinguish several locations where the beam was scanning that have become cleaner in large square patches. In (b, c) we can clearly see the graphene lattice, illustrating how clean it becomes and the interesting adsorbed atoms on the surface, in holes, and substitutional positions. (d, e) are the first and last frames from the accompanying time lapse series (3.7 minutes real time) of the contamination dynamics.

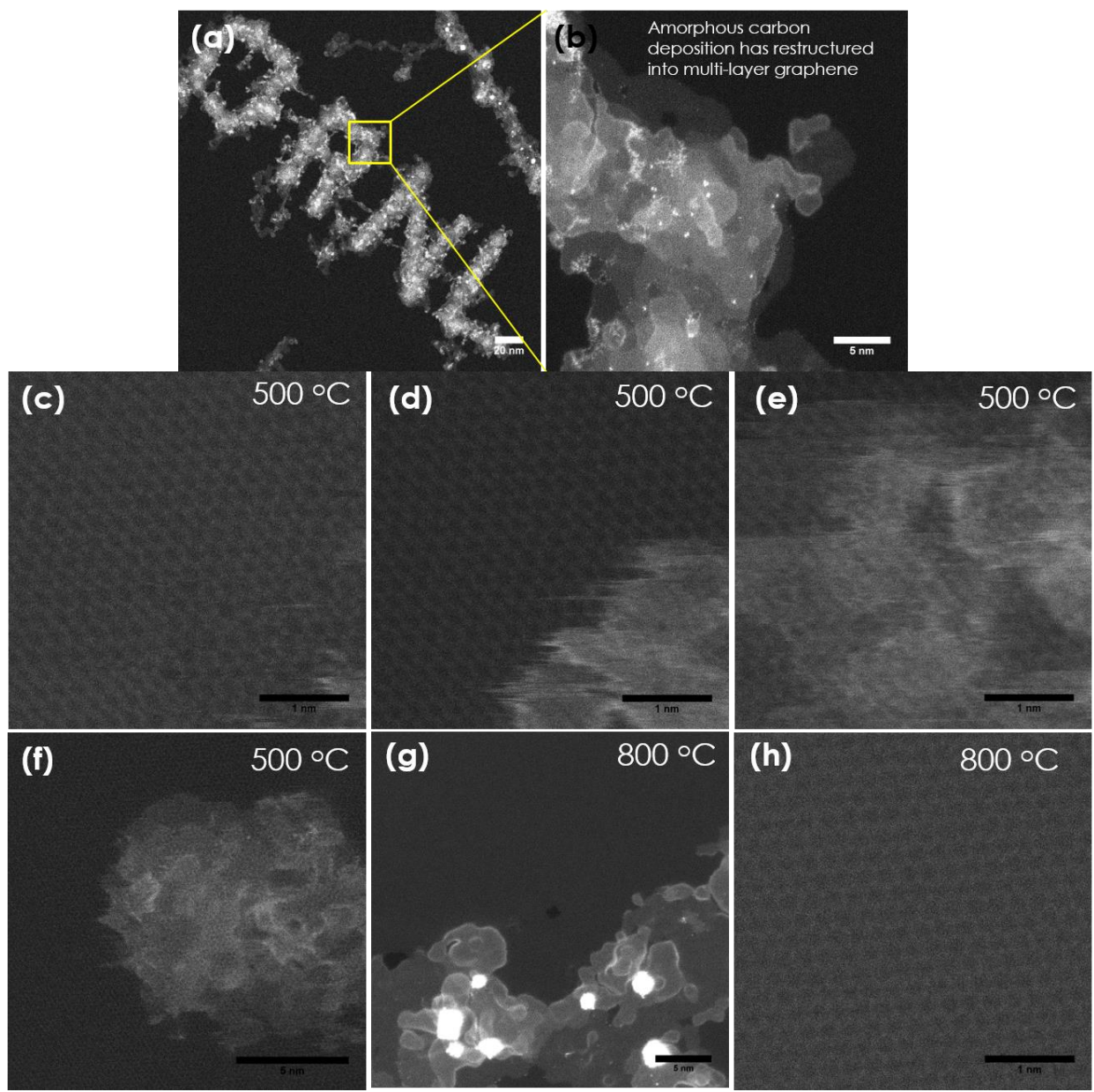

**Figure S2** (a, b) show the conversion of the amorphous hydrocarbon deposition from Figure 4 (b) in the text to graphitic carbon upon heating to 1200 °C. This is evidenced by the discrete steps in intensity indicating steps in number of layers of the underlying material which is not present in amorphous materials. (c, d, e) show a series of images acquired back-to-back at 500 °C after the initial 1200 °C cleaning. We observe a steady deposition of hydrocarbon contamination. Rather than amorphous carbon, this appears to be forming a layered structure indicative of graphite. (f) shows a lower magnification view of the deposition introduced in (c, d, e) where the intensity steps are more clearly discernable. (g, h) are images acquired at 800 °C after the initial 1200 °C cleaning. E-beam deposition was no longer observed.

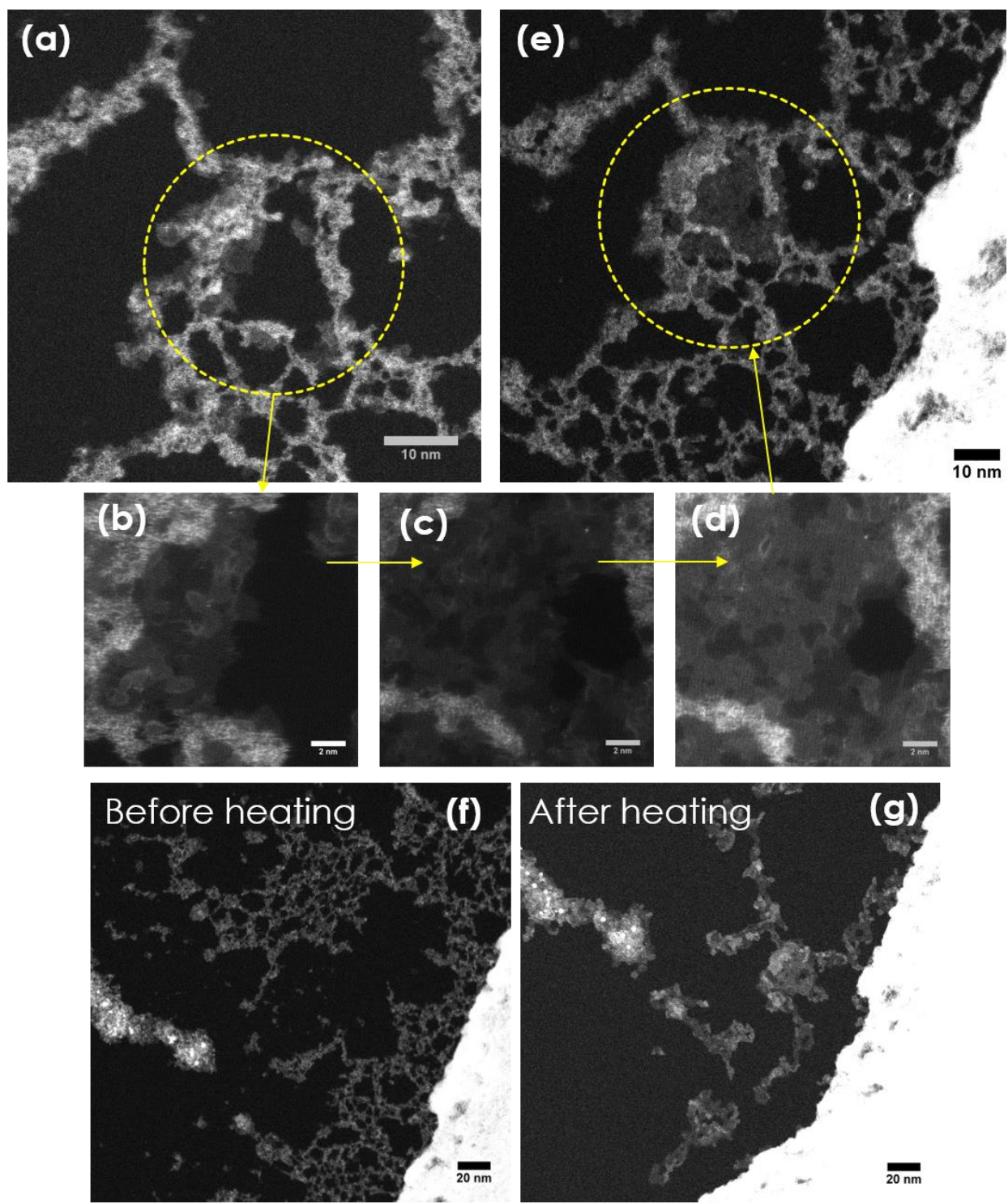

**Figure S3** (a, b, c, d, e) illustrate the growth of graphitic carbon contamination under e-beam exposure. (b, c, d) are images scanned over the area circled in (a) and the circle in (e) shows the result. (f, g) show the dendritic structure before and after heating to 1200 ºC. We see that most of this contamination is freed by the heating process but some remained stuck to the surface. It is unclear if this is due to e-beam exposure prior to heating.